\documentclass[letterpaper, 10 pt, conference]{ieeeconf}
\IEEEoverridecommandlockouts
\let\labelindent\relax
\usepackage{enumitem}
\usepackage{balance}
\usepackage[font={small}]{caption}
\usepackage{subcaption}
\usepackage{array}
\usepackage{textcomp}
\usepackage{mathtools, nccmath}
\usepackage{graphicx}
\usepackage{amsfonts}
\usepackage{amsmath}
\usepackage{amssymb}
\usepackage{algorithm}
\usepackage{algorithmic}
\usepackage{hyperref}
\usepackage{tikz}
\usepackage{arydshln}
\usepackage{multirow}
\usepackage{bm}
\usepackage{epstopdf}
\usepackage{cite}
\usepackage{siunitx}

\DeclareMathOperator*{\argmin}{argmin} 

\newtheorem{remark}{Remark}

\newcommand{\he}[1]{{\color{cyan} #1}}
\begin{document}

\title{\LARGE \bf Rule-Based Safety-Critical Control Design using Control Barrier Functions with Application to Autonomous Lane Change}

\author{Suiyi He, Jun Zeng, Bike Zhang and Koushil Sreenath
        \thanks{All authors are with the University of California, Berkeley, CA, 94720, USA, \tt{\small \{suiyi\_he, zengjunsjtu, bikezhang, koushils\}@berkeley.edu} }
        \thanks{This work was partially supported through National Science Foundation Grant CMMI-1931853.}
        \thanks{Code is available at \url{https://github.com/HybridRobotics/Lane-Change-CBF}.}
        \thanks{Simulation videos are at \url{https://youtu.be/icmy9u2a4z4}.}
}

\maketitle
\begin{abstract}
This paper develops a new control design for guaranteeing a vehicle's safety during lane change maneuvers in a complex traffic environment. The proposed method uses a finite state machine (FSM), where a quadratic program based optimization problem using control Lyapunov functions and control barrier functions (CLF-CBF-QP) is used to calculate the system's optimal inputs via rule-based control strategies. The FSM can make switches between different states automatically according to the command of driver and traffic environment, which makes the ego vehicle find a safe opportunity to do a collision-free lane change maneuver. By using a convex quadratic program, the controller can guarantee the system's safety at a high update frequency. A set of pre-designed typical lane change scenarios as well as randomly generated driving scenarios are simulated to show the performance of our controller. 
\end{abstract}
\IEEEpeerreviewmaketitle
\section{Introduction}
\subsection{Motivation}

\he{}
Safety is critical in the automobile industry, and significant progress in this area has been made via (semi-)autonomous driving in the past few decades. While systems like adaptive cruise control or lane departure warning systems have proven to improve vehicle's safety in simple scenarios, it is desirable in the future that autonomous driving could handle safety-critical tasks in more complex driving scenarios, for example, autonomous lane change maneuvers. Lane change is challenging for both drivers and autonomous driving controllers since multiple surrounding vehicles should be considered and their future movements should be predicted. The more challenging fact is that drivers or controllers have little time to respond to avoid a crash if a threat occurs. According to statistics provided by US National Highway Traffic Safety Administration (NHTSA) in \cite{sen2003analysis}, about 9\% of all vehicle crashes involved lane change maneuvers.   
\vspace{-2mm}
\subsection{Related Work}
Several recent work focuses on this topic. In \cite{wissing2017probabilistic,hu2018probabilistic}, new methods are proposed to predict the intent of other surrounding vehicles. Research in \cite{werling2010optimal,schmidt2019interaction} focuses on the generation of optimal trajectories for lane change maneuvers. Various low-level controllers have been developed to track the optimal path generated by planners. Model predictive control (MPC) is a commonly used method. In \cite{bae2020cooperation,zheng2013model,chen2019lane}, MPC based controllers are implemented. Additionally, simple linear \cite{joshi2020autonomous} or nonlinear \cite{chen2020impaired} feedback control method has also been proven to be suitable for this task.
\begin{figure}[t]
    \setlength{\abovecaptionskip}{0.5cm}
    \setlength{\belowcaptionskip}{-0.5cm}
    \centering
    \includegraphics[width=1.0\linewidth]{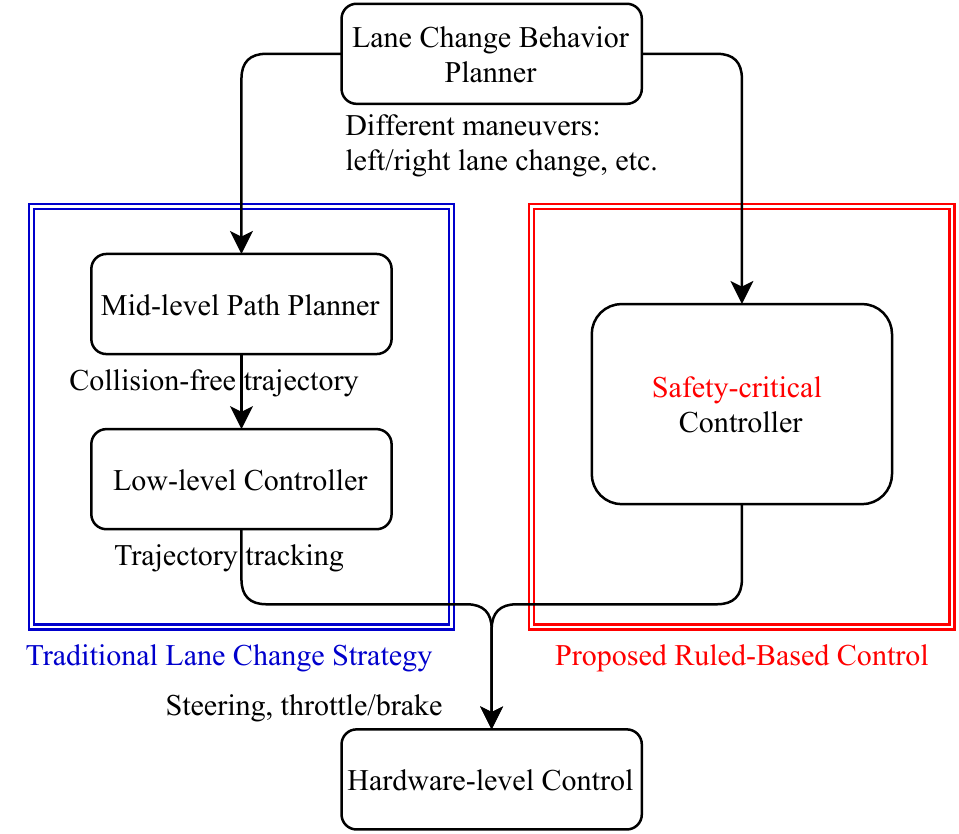}
    \caption{In our proposed lane change strategy, the mid-level path planner and low-level controller in the traditional strategy are unified into one optimization problem. This improves the computational efficiency, which allows the strategy to guarantee the system's safety in fast changing environments through a low-level safety-critical planner-controller.}
    \label{fig:lane-change-strategy}
\end{figure}

Based on previous work like \cite{zheng2013model,chen2019lane}, we can summarize the traditional autonomous lane change strategy as shown in the left side of Fig. \ref{fig:lane-change-strategy}. A high-level behavior planner makes the decision whether the ego vehicle should stay in the current lane or change to the left (right) adjacent lane. This can be done by human drivers (e.g. activating the turn signal) or by algorithms \cite{vallon2017machine}. Once a specific lane change maneuver is triggered, a mid-level path planner generates a collision-free optimal trajectory for the ego vehicle, which will be tracked by a low-level controller. However, the mid-level path planner usually can not be updated at a high frequency and may not respond to a sudden threat immediately, which means its planned trajectory may not always meet safety-critical requirements in a fast changing environment. Furthermore, there are always notable tracking errors in the low-level controller, which means tracking a collision-free trajectory could still be potentially unsafe. Therefore, in a safety-critical autonomous lane change strategy, safety relevant constraints should be considered in the low-level controller directly and the controller must work at a high update frequency.

In order to guarantee the system's safety in the low-level controller, control barrier functions (CBFs) have recently been introduced to ensure set invariance by considering the system dynamics and several researchers have used CBFs to ensure the system's safety for the vehicle control design problems. In \cite{chen2017obstacle, son2019safety, zeng2020safety, rosolia2020multi}, CBFs are used to implement obstacle avoidance. A CBF-based controller to supervise the safety of a learning based lane keeping controller is proposed in \cite{chen2019enhancing}. In \cite{ames2014control}, CBFs are unified with control Lyapunov functions (CLFs) via a quadratic program, and it illustrates this CLF-CBF-QP formulation in the context of adaptive cruise control. Due to its low computational complexity, this quadratic program allows the low-level controller to work at a high update frequency, which motivates us to investigate the safety-critical lane change control design through the CLF-CBF-QP formulation.  

Rule-based strategies are also widely used in control design. One example is the finite state machine (FSM), which can make decisions according to input signals and transition conditions \cite{montemerlo2008junior}. Since its computational complexity is lower than that of a traditional path planner, the FSM can be used in a low-level controller and work at a high update frequency, which inspires us to use a well-designed FSM to replace the mid-level path planner in traditional lane change strategies. 

Based on the above analysis, we propose a rule-based safety-critical lane change control design. A FSM works as the basic structure, where a quadratic program based optimization problem using CLF and CBF constraints (CLF-CBF-QP) is formulated to achieve control objectives and guarantee the system's safety. The FSM can unify the mid-level path planner and low-level controller in traditional lane change strategy as one: a low-level safety-critical controller, see Fig. \ref{fig:lane-change-strategy}. The FSM is used to make the decision if the ego vehicle can do a collision-free lane change maneuver or not. If the current traffic environment isn't suitable for a lane change maneuver, the ego vehicle will keep the current lane until a safe situation occurs. Additionally, if a threat arises during a lane change maneuver, the FSM will enter another state to drive the ego vehicle back to its current lane. Since the safety relevant constraints are considered in the low-level controller directly and a quadratic program allows the controller to run with high update frequency, our proposed strategy can guarantee the ego vehicle's safety in complex environments.
\vspace{-2mm}

\subsection{Contribution}
The contribution of this paper is as follows:
\begin{itemize}
    \item We present a rule-based safety-critical design by using CLF-CBF-QP formulation, which can achieve the control objective and guarantee the vehicle's safety via low-level control in lane change maneuvers. The quadratic program allows the controller to work at a high update frequency.
   
    \item A finite state machine is used to unify mid-level planner and low-level controller into one optimization problem, a low-level safety-critical controller. The FSM helps the proposed strategy do switches between multiple CBF constraints in a complicated environment. 

    \item We verify the performance of our controller using both typical and randomly generated driving scenarios. The result shows that our approach is generally applicable in both highway and city driving scenarios.
\end{itemize}
\vspace{-2mm}
\subsection{Paper Structure}
This paper is organized as follows:
in Sec. \ref{Sec:Background}, we present the background of vehicle model and optimal control with control barrier functions and control Lyapunov functions.
In Sec. \ref{Sec:Control-Design}, the safety-critical controller for lane change maneuver is illustrated.
To validate the performance of our control design, pre-designed typical scenarios and randomly generated cases are used to test the controller in Sec. \ref{Sec:Results}.
Sec. \ref{Sec:Discussion} analyzes the controller and shows the potential future work.
Sec. \ref{Sec:Conclusion} presents concluding remarks.
\section{Background}
\label{Sec:Background}
\subsection{Vehicle Model}

In this paper, we use a kinematic bicycle model \eqref{eq:kinematic-model} in \cite{Kong2015} for our numerical validation and its dynamics is described as follows,
\begin{subequations}
\label{eq:kinematic-model}
\begin{align}
\Dot{x} \quad &= \quad v  \cos{(\psi + \beta)} \label{Xa} \vspace{1ex} \\
\Dot{y} \quad &=\quad v  \sin{(\psi + \beta)}  \vspace{1ex}\\
\Dot{\psi} \quad &=\quad  \dfrac{v}{l_r}\sin{\beta} \vspace{1ex}\\
\Dot{v} \quad &= \quad a \vspace{1ex} \label{Xd} \\
\beta \quad &=\quad \tan^{-1} \left(\dfrac{l_r}{l_f+l_r}\tan \left(\delta_f \right) \right) \label{Xe}
\end{align}
\end{subequations}
where acceleration at vehicle's center of gravity (c.g.). $a$ and front steering angle $\delta_f$ are the inputs of the system. $x$ and $y$ denote the coordinates of the vehicle's c.g. in an inertial frame (\textit{X},\textit{Y}). $\psi$ represents the orientation of the vehicle. $l_f$ and $l_r$ describe the distance from vehicle's c.g. to the front and real axles, respectively. $\beta$ represents the slip angle of the vehicle. 
\begin{remark}
\label{rem:kinematic-dynamic-model-limitations}
    Both kinematic and dynamic bicycle models are commonly used vehicle models in the field of autonomous driving research.
    From \cite{polack2017kinematic}, authors show that the kinematic bicycle model works under small lateral acceleration and
    also recommend 0.5$\mu g$ as a limitation for the validity of the kinematic bicycle model ($\mu$ is friction coefficient and $g$ is gravitational acceleration).
    In this work, we assume our controller will result in small longitudinal acceleration and this criteria will be considered as a constraint in our optimal control problem in Sec. \ref{Sec:Control-Design}
\end{remark}

Notice that the kinematic model in \eqref{eq:kinematic-model} represents a nonlinear nonaffine system. For our later usage, we assume that the slip angle $\beta$ is constrained with a small angle assumption in our control design together, where we approximate $\cos\beta=1$ and $\sin\beta=\beta$. 
Hence, \eqref{eq:kinematic-model} could be simplified as a nonlinear affine form as follows,

\begin{equation}
\label{eq:affine-kinematic-model}
\begin{bmatrix}
\dot{x} \\ \dot{y} \\ \dot{\psi} \\ \dot{v}
\end{bmatrix} =
\begin{bmatrix}
v\cos\psi \\ v\sin\psi \\ 0 \\ 0
\end{bmatrix} +
\begin{bmatrix}
0 & -v\sin\psi \\
0 & v\cos\psi \\
0 & v/l_r \\
1 & 0
\end{bmatrix}
\begin{bmatrix}
a \\ \beta
\end{bmatrix}
\end{equation}

This nonlinear affine model in \eqref{eq:affine-kinematic-model} with inputs $a$ and $\beta$ will be used for our safety-critical control design. Input $\delta_f$ in \eqref{eq:kinematic-model} can be calculated through \eqref{Xe}

\subsection{Safety-Critical Control}
Consider a nonlinear affine control system
\begin{equation}
    \dot{\mathbf{x}} = f(\mathbf{x}) + g(\mathbf{x}) \mathbf{u}, \label{eq:standard-nonlinear-affine-system}
\end{equation}
where $\mathbf{x} \in D \subset \mathbb{R}^n$, $\mathbf{u} \in \mathcal{U} \subset \mathbb{R}^m$ represent the system's state and input, and $\mathcal{U}$ is the admissible input set of the system, $f$ and $g$ are locally Lipschitz.

For this nonlinear affine system, we are interested in the system's safety. For the same control system \eqref{eq:standard-nonlinear-affine-system}, we define a superlevel set $\mathcal{C}\subset D \subset \mathbb{R}^n$ of a differentiable function $h:$ $ D \times \mathbb{R} \to \mathbb{R}$,
\begin{equation}
    \mathcal{C}=\left\{ \mathbf{x} \in \mathbb{R}^n : h(\mathbf{x}, t) \geq 0 \right\},
\end{equation}
and we refer to $\mathcal{C}$ as a safe set. The function $h$ becomes a control barrier function (CBF) if it can satisfy the following condition \cite{wu2016safety}:
\begin{equation}
    \sup\limits_{\mathbf{u} \in \mathcal{U}} [\dfrac{\partial h}{\partial t} + L_f h(\mathbf{x}, t) + L_g h(\mathbf{x}, t)\mathbf{u}] \geq -\gamma h(\mathbf{x}, t). \label{eq:time-varying-cbf-definition}
\end{equation}
When $h(\mathbf{x},t)$ is time-invariant, denoted as $h(\mathbf{x})$, then the condition above could be simplified \cite{ames2019control} as follows, 
\begin{equation}
    \sup\limits_{\mathbf{u} \in \mathcal{U}} [L_f h(\mathbf{x}) + L_g h(\mathbf{x})\mathbf{u}] \geq -\gamma h(\mathbf{x}). \label{eq:cbf-definition}
\end{equation}
The control barrier function guarantees the set invariance of $\mathcal{C}$ for the system's safety. Besides the system's safety, we have a control Lyapunov function for the system's stability, where a function $V : D \in \mathbb{R} ^ n \to \mathbb{R} _{+}$ is a control Lyapunov function (CLF) if (i) V is positive definite and (ii) it can satisfy the following condition:
\begin{equation}
    \inf\limits_{\mathbf{u} \in \mathcal{U}} [L_f V(\mathbf{x})+L_g V(\mathbf{x})\mathbf{u} \leq -\alpha V(\mathbf{x})].
    \label{eq:clf-definition}
\end{equation}
Notice that $\gamma$ and $\alpha$ in \eqref{eq:time-varying-cbf-definition} and \eqref{eq:clf-definition} could be generalized into extended class $\mathcal{K}_{\infty}$ and $\mathcal{K}$ functions, respectively; and we only treat them as linear functions with constant coefficients in this paper.

To guarantee the system's safety and achieve its control objective simultaneously, the control Lyapunov function and control barrier function can be unified as a quadratic program (CLF-CBF-QP) \cite{ames2014control} and written as follows:
\begin{subequations}
\label{eq:clf-cbf-qp}
\begin{align}
    \mathbf{u}(\mathbf{x}) &= \argmin_{(\mathbf{u},\delta) \in \mathbb{R}^{m+1} } \dfrac{1}{2}\mathbf{u}^TH(\mathbf{x})\mathbf{u}+p\delta^2 \\
    \text{s.t.} & \quad L_fV(\mathbf{x})+L_gV(\mathbf{x})\mathbf{u}\leq -\alpha V(\mathbf{x}) +\delta \\
    & \quad \dfrac{\partial h}{\partial t}+ L_f h(\mathbf{x}, t) + L_g h(\mathbf{x}, t) \mathbf{u} \geq -\gamma h(\mathbf{x}, t) 
\end{align}
\end{subequations}
where $H(\mathbf{x})$ is a positive definite matrix, and $\delta$ is a relaxation variable to make the CLF constraint become a soft constraint to mediate stability for safety, and $p$ is the penalty for this relaxation variable. This formulation will be used for our control design, which will be introduced in Sec. \ref{Sec:Control-Design}.
\section{Control Design} \label{Sec:Control-Design}
After introducing the vehicle model and CLF-CBF-QP formulation, a safety-critical lane change controller will be presented in this section. The proposed controller is a finite state machine (FSM), where rule-based CLF-CBF-QP formulations are used to calculate the system's optimal input. This FSM will be introduced in Sec. \ref{sec:finite-state-machine}. Then we will present the safety-based conditions for switches of constraints in Sec. \ref{sec:safety-condition-switch-cbf}. Finally, details about the CLF-CBF-QP formulations will be shown in Sec. \ref{sec:Rule-based-strategy}.  

\begin{figure}
    \setlength{\abovecaptionskip}{0.5cm}
    \setlength{\belowcaptionskip}{-0.5cm}
    \centering
    \includegraphics[width=\linewidth]{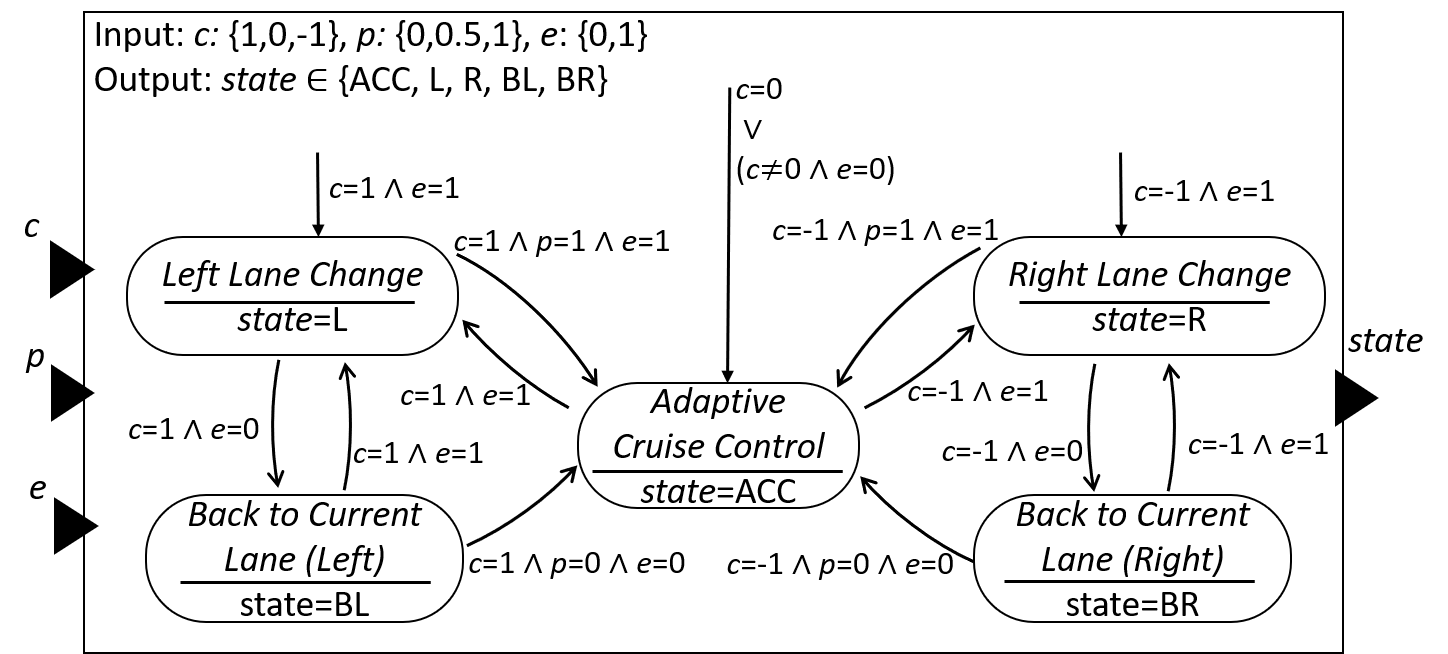}
    \caption{Finite state machine of lane change controller. The command from behaviour planner ($c$), current traffic environment ($e$) and the ego vehicle's positional information ($p$) work as the input signals to this finite state machine and its output is the controller's state. }
    \label{fig:finite-state-machine}
\end{figure}
\subsection{Finite State Machine} \label{sec:finite-state-machine}
Fig. \ref{fig:finite-state-machine} shows the proposed FSM. The FSM's output is a state representing one of the following:

\noindent\textit{Adaptive Cruise Control State} - \textbf{ACC}: The ego vehicle maintains a desired speed and follows a leading vehicle in its current lane at a safe distance.

\noindent\textit{Left or Right Lane Change State} - \textbf{L} or \textbf{R}: The ego vehicle is expected to do a collision-free lane change maneuver to the left or right adjacent lane, respectively.

\noindent\textit{Back to Current Lane From Left or Right State} - \textbf{BL} or \textbf{BR}: The ego vehicle drives back to its current lane to avoid a potential crash if a threat arises during a lane change maneuver.

Having presenting the FSM's state definitions, we next introduce the FSM's input signals ($c, p, e$) and reveal how they make the FSM switch between the above states.

\noindent \textit{Command from High-Level Behaviour Planner ($c$)} : This indicates the high-level planner's expected maneuver for the ego vehicle. Value 0 will set the controller in \textbf{ACC} state; value 1 or -1 will make the controller work in \textbf{L} or \textbf{R} state, respectively.

\noindent \textit{Positional Information - ($p$)} : This represents the ego vehicle's relative lateral position. Value 0 means the ego vehicle is in its current lane; if it moves across the edge between current and target lanes, $p$ will change to 0.5; finally, when the ego vehicle is totally in its target lane for more than some duration of time, e.g. 1.5s, $p$ will become 1, which represents the success of a lane change maneuver and will bring the controller back to \textbf{ACC} state.

\noindent \textit{Traffic Environment Information - ($e$)} : This shows whether the ego vehicle can do a lane change maneuver under safety-critical constraints. When the CLF-CBF-QP formulation is in the \textbf{L} or \textbf{R} state and is numerically unsolvable due to a potential future collision, $e$ will change from 1 to 0. When $c$ is not 0 but $e$ is 0, then if the FSM is in \textbf{ACC} state, it will continue working in this state; otherwise, as shown in Fig. \ref{fig:finite-state-machine}, the FSM will go back to \textbf{ACC} state via \textbf{BL} or \textbf{BR} state.

When $c$'s value is -1 or 1 but the ego vehicle is in \textbf{ACC} state, a predictive calculation as in \eqref{eq:predictive-calculation} will be made to determine if the ego vehicle can get enough space for a lane change maneuver after accelerating to the speed limit. When the result shows this is possible, the speed limit will be the desired speed for \textbf{ACC} state and the ego vehicle will re-enter \textbf{L} or \textbf{R} state once the lane change CLF-CBF-QP is solvable.

These switches between different states implement the function of a planner in our proposed strategy. According to the input signals, the FSM will decide when is the best opportunity to do the lane change maneuver. Additionally, if this maneuver is interrupted, the FSM will drive the ego vehicle to change to its target lane again once it is safe. 

\subsection{Safety-Based Conditions for Switches of CBFs } \label{sec:safety-condition-switch-cbf}
\begin{figure}
    \setlength{\abovecaptionskip}{0.5cm}
    \setlength{\belowcaptionskip}{-0.5cm}
    \centering
    \includegraphics[width=\linewidth]{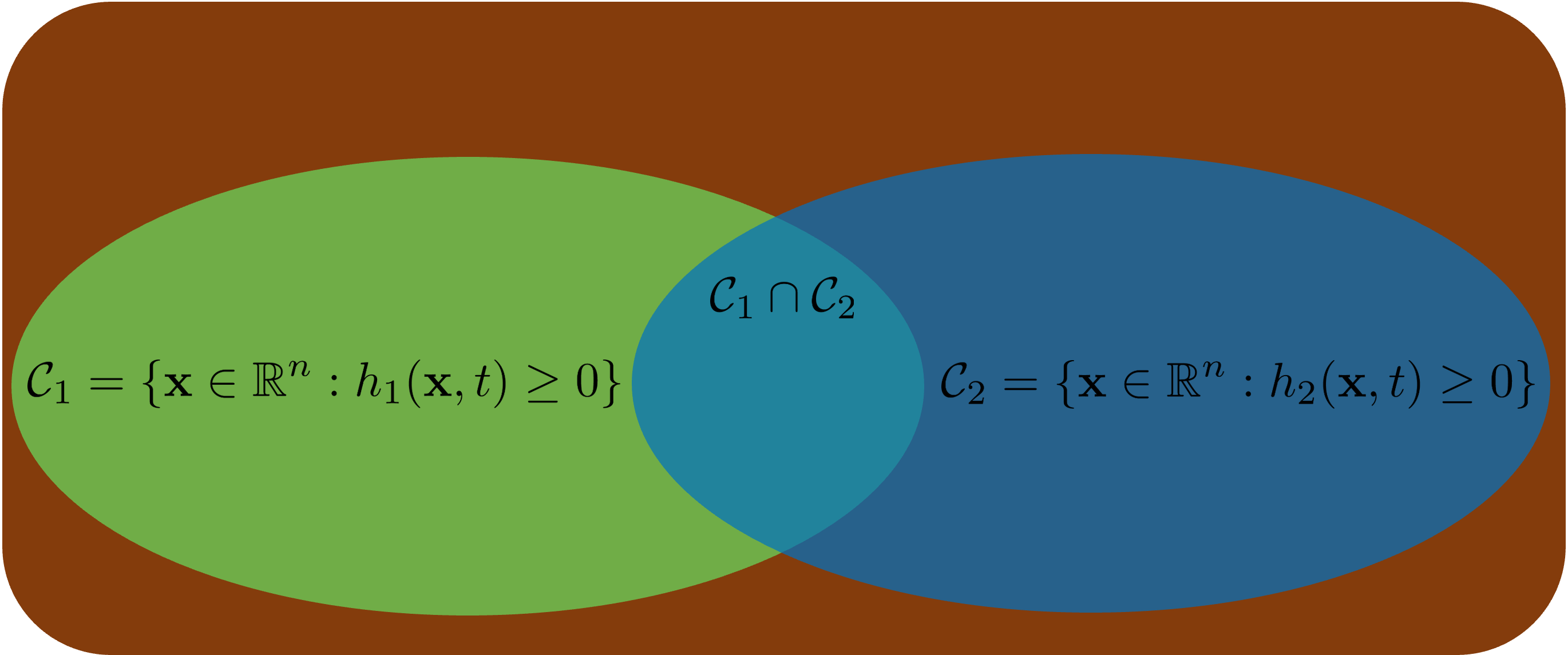}
    \caption{Safe sets' invariance for switches of CBF constraints. The green region and blue region represent two safe sets, respectively. The brown region shows the unsafe set. When the controller does a switch between two safe sets, the system's state $\mathbf{x}$ must be in the intersection of two safe sets.}
    \label{fig:CBF-switch-condition}
\end{figure}

In this section, we will show the continuity of the system's safety under different CBF constraints. In our controller, the switches between different CBF constraints could happen when the FSM changes state or a CBF constraint is removed from the CLF-CBF-QP formulation (see Sec. \ref{sec:Rule-based-strategy}). This asks the controller to guarantee the continuity of system's safety. As shown in Fig. \ref{fig:CBF-switch-condition}, system's CBF constraints before and after a switch can be described as safe sets $\mathcal{C}_1$ and $\mathcal{C}_2$, respectively. When $h_2(\mathbf{x},t)$ replaces $h_1(\mathbf{x},t)$ as the new CBF constraint, if the system's state $\mathbf{x}$ is in the intersection of $\mathcal{C}_1$ and $\mathcal{C}_2$, the new barrier function $h_2(\mathbf{x},t)$ will make the new safe set $\mathcal{C}_2$ invariant after the switch, which will guarantee the system's safety. This condition is used to design switches between different CBF constraints in this paper.

\subsection{Rule-Based CLF-CBF-QP Formulation} \label{sec:Rule-based-strategy}
\begin{figure}
    \centering 
    \includegraphics[width=\linewidth]{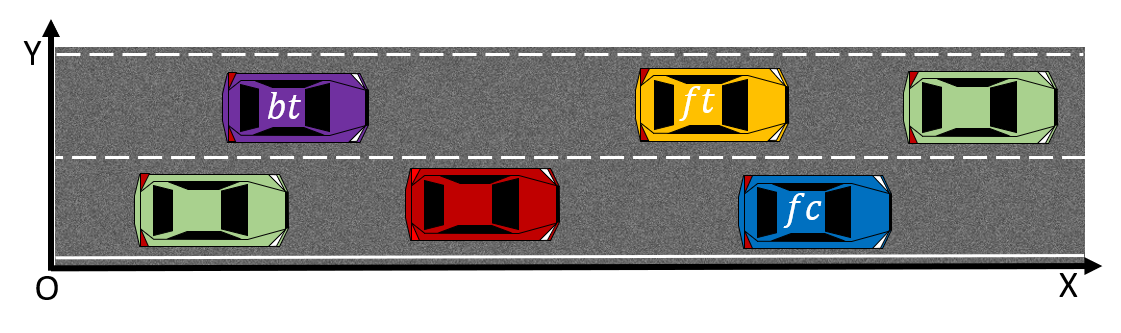}
    \caption{A typical lane change scenario. Red indicates the ego vehicle; vehicles with letters are safety-critical relevant vehicles. Green vehicles will not be considered in control design. The frame is the inertial frame $E$ that is used in further discussion.}
    \label{fig:normal-lane-change}
\end{figure}

We now present the CLF-CBF-QP formulations in different FSM states. Firstly, we consider a typical lane change scenario as shown in Fig. \ref{fig:normal-lane-change}, where the red vehicle is our ego vehicle. In the controller, up to three vehicles will be selected as vehicles of interest (vehicles with letters in Fig. \ref{fig:normal-lane-change}): vehicle $fc$ represents the vehicle immediately in \emph{front} of the ego vehicle in its \emph{current lane} (blue vehicle in Fig. \ref{fig:normal-lane-change}); vehicle $bt$ represents the vehicle immediately \emph{behind} the ego vehicle in the \emph{target lane} (purple vehicle in Fig. \ref{fig:normal-lane-change}); vehicle $ft$ denotes the vehicle immediately in \emph{front} of the ego vehicle in the \emph{target lane} (yellow vehicle in Fig. \ref{fig:normal-lane-change}). Other surrounding vehicles like green ones in the Fig. \ref{fig:normal-lane-change} will not be considered in the controller. The frame in Fig. \ref{fig:normal-lane-change} is the inertial frame used in further discussion, which is called $E$. 

\begin{remark}
In this work, we are interested in the safety-critical control of the autonomous lane change maneuver. To explore this problem, we assume the controller has access to accurate data of all surrounding vehicles. This could be done through Lidar, vision, radar, and ultrasonic sensors.  
\end{remark}

TABLE \ref{tab:notation-for-control-design} shows notations that will be used for further discussion and the subscript $k$ can represent $fc$, $bt$ or $ft$, which indicates the corresponding vehicle with respect to the ego vehicle.
\begin{table}
    \centering
    \caption{Notations and Symbols for control design.}
    \begin{tabular}{c c}\hline
        Notation & Description \\ \hline
        \multicolumn{2}{c}{\textbf{Vehicle's Dimension Data}}\\
        $l_{fc}$& length of vehicle's body that is in front of the c.g.\\
        $l_{rc}$& length of vehicle's body that is behind the c.g.\\
        $w_{lc}$& width of vehicle's body that is on the left of c.g.\\
        $w_{rc}$&width of vehicle's body that is on the right of c.g.\\\hline
        \multicolumn{2}{c}{\textbf{Ego Vehicle's Data}}\\
        $(x,y)$&coordinates of c.g. in frame $E$ \\
        $v$& ego vehicle's speed \\
        $\psi$& ego vehicle's yaw angle \\
        $v_{d}$ & ego vehicle's desired speed \\
        $v_{l}$ & ego vehicle's speed limit of current scenario\\
        $a_{l}$ & ego vehicle's acceleration limit\\
        $\epsilon$& a safety factor between 0.1-1 (1 is safest)\\\hline
        \multicolumn{2}{c}{\textbf{Other Vehicles' Data}}\\
        $(x_k,y_k)$&coordinates of vehicle $k$'s c.g. in frame $E$ \\
        $v_k$ & vehicle $k$'s speed \\
        $a_k$ & vehicle $k$'s acceleration \\
        \multirow{2}{*}{$\Delta x_k$}& time varying longitudinal distance between vehicle $k$ \\~&and the ego vehicle, equal to $|x-x_k|-l_{fc}-l_{rc}$. \\
        \multirow{2}{*}{$\Delta y_k$}&time varying lateral distance between vehicle $k$ and \\~&the ego vehicle, equal to $|y-y_k|-w_{rc}-w_{lc}$.\\\hline
    \end{tabular}
    \label{tab:notation-for-control-design}
\end{table}

\begin{remark}
The vehicle's position, velocity or yaw angle information is a function of time. In order to simplify the notations for discussion, in this paper, we omit time information in the notations. For example, $v$ represents the speed of the ego vehicle at the current time.  
\end{remark}

We next present constraints used in the optimization problem, divided into two groups:

\noindent\textbf{Hard Constraints:} These represent the safety-critical relevant constraints. In our controller, these constraints are used to keep the ego vehicle at a safe distance from the surrounding vehicles, which is defined by us as 1+$\epsilon$ times following vehicle's speed. For example, if the ego vehicle changes its lane, the distance between it and the vehicle $bt$ should be greater than (1+$\epsilon$)$v_{bt}$. Hard constraints should never be violated under any conditions and will be guaranteed through CBF constraints.

\noindent\textbf{Soft Constraints:} These introduce the control objectives related constraints. The control goals will only be achieved through CLFs when the hard constraints are satisfied, for example the speed or position of the ego vehicle. Through CLF constraints, it is possible to track a desired position without a reference trajectory.

Next we present details of the CLF-CBF-QP formulations in each FSM state.

For all FSM states, the following CLFs will be used to regulate the ego vehicle to track the desired speed $v_d$ and reach its lateral dynamics' control objective:

\begin{subequations}\label{eq:CLF}
    \begin{align}
            V_{v}(\mathbf{x})&=(v-v_{d})^2,\\ 
            V_{y}(\mathbf{x})&=(y-y_{l})^2,\\
            V_{\psi}(\mathbf{x})&=\psi^2,
    \end{align}
\end{subequations}
where $y_l$ is the y coordinate of current lane's center line for states \textbf{ACC}, \textbf{BL} and \textbf{BR} or y coordinate of target lane's center line for states \textbf{L} and \textbf{R} in frame $E$. 

As mentioned in Sec. \ref{sec:finite-state-machine}, if input $c$ is 1 or -1 but the FSM is in \textbf{ACC} state, a simple predictive calculation will be done to determine the ego vehicle's desired speed by first computing the distances between the ego vehicle and the three vehicles of interest:
\begin{small}
\begin{subequations}
\begin{align}
    \Delta x_{fc}'&=\Delta x_{fc}+v_{fc}\dfrac{v_{l}-v}{a_{l}}-\dfrac{v_{l}^2-v^2}{2a_{l}}-(1+\epsilon)v,\\
    \Delta x_{ft}'&=\Delta x_{ft}+v_{ft}\dfrac{v_{l}-v}{a_{l}}-\dfrac{v_{l}^2-v^2}{2a_{l}}-(1+\epsilon)v,\\
    \Delta x_{bt}'&=\Delta x_{bt}-v_{bt}\dfrac{v_{l}-v}{a_{l}}+\dfrac{v_{l}^2-v^2}{2a_{l}}-(1+\epsilon)v_{bt}.
\end{align}
\label{eq:predictive-calculation}
\end{subequations}
\end{small}
If all equations above are greater than 0, which means the ego vehicle will have enough space to change the lane through accelerating to the current scenario's speed limit $v_l$, $v_l$ will become the new desired speed $v_d$ in \textbf{ACC} state; otherwise, the original desired speed $v_d$ will be used in \textbf{ACC} state. 

\noindent\textbf{ CBF in ACC state}: In this state, safety-critical control should keep the distance between the ego vehicle and vehicle $fc$ greater than a pre-defined value. We refer to the distance constraints and force based constraints in \cite{ames2014control} and construct the following CBF:

\begin{small}
\begin{equation}
h_{fc}(\mathbf{x},t){=}\bigg\{\begin{array}{ll}
     \Delta x_{fc}{-}(1+\epsilon) v{-}\dfrac{(v_{fc}-v)^2}{2a_{l}}& \textrm{if $v\geq v_{fc}$}  \\
    \Delta x_{fc}{-}(1+\epsilon) v & \textrm{else}
\end{array}
\label{eq:hfc}
\end{equation}
\end{small}

If the ego vehicle is faster than its leading vehicle $fc$, the traveling distance during deceleration process will be considered in the $h_{fc}(\mathbf{x},t)$. Otherwise, $h_{fc}(\mathbf{x},t)\geq 0$ indicates the ego vehicle meets the safety-critical requirement directly. Converting this CBF into its corresponding constraint in the CLF-CBF-QP formulation will guarantee $h_{fc}(\mathbf{x},t)$ always greater than 0. This condition will also be used for similar expressions later.

\noindent\textbf{ CBFs in L or R state}:
In this state, all three vehicles of interest should be considered in the safety-critical control design. Following CBFs will be constructed:
{\small
\begin{subequations}
    \begin{align}
        h_{fc}(\mathbf{x},t)&{=}\bigg\{\begin{array}{ll}
        \Delta x_{fc}{-}(1+\epsilon) v{-}\dfrac{(v_{fc}-v)^2}{2a_{l}}& \textrm{if $v\geq v_{fc}$}  \\
        \Delta x_{fc}{-}(1+\epsilon) v & \textrm{else}
        \label{eq:l-fc}
        \end{array} \\
        h_{ft}(\mathbf{x},t)&{=}\bigg\{\begin{array}{ll}
        \Delta x_{ft}{-}(1+\epsilon) v{-}\dfrac{(v_{ft}-v)^2}{2a_{l}}& \textrm{if $v\geq v_{ft}$}  \\
        \Delta x_{ft}{-}(1+\epsilon) v & \textrm{else}
        \label{eq:l-ft}
        \end{array} \\
        h_{bt}(\mathbf{x},t)&{=}\bigg\{\begin{array}{ll}
        \Delta x_{bt}{-}(1+\epsilon) v_{bt}{-}\dfrac{(v_{bt}{-}v)^2}{2a_{l}}&\textrm{if $v_{bt}\geq v$}   \\
        \Delta x_{bt}{-}(1+\epsilon) v_{bt} & \textrm{else}
        \label{eq:hbt}
        \end{array}
    \end{align}
\end{subequations}
}

Similarly to the \eqref{eq:hfc}, \eqref{eq:hbt} can be used to keep a safe distance between the ego vehicle and vehicle $bt$ (or after the ego vehicle accelerates to the same speed as vehicle $bt$). Additionally, during a lane change maneuver, after the ego vehicle changes to its target lane, vehicle $fc$ and $bt$ will no longer be the vehicles of interest. Therefore, $h_{fc}(\mathbf{x},t)$ and $h_{bt}(\mathbf{x},t)$ will not be used if the ego vehicle is totally in its target lane.

\textbf{CBFs in BL or BR state}:
In this state, since the ego vehicle will go back to its current lane, it should keep a safe distance from vehicle $fc$ by using  the CBF as in \eqref{eq:hfc-blbr}. More importantly, hard constraints should be used to prevent a potential crash with interrupted vehicles, which can be either vehicle $ft$ or $bt$. Equations \eqref{eq:hft-blbr} and \eqref{eq:hbt-blbr} are used to prevent a crash in both longitudinal and lateral directions (if the vehicle $ft$ or $bt$ is overlapping with the ego vehicle longitudinally). Following CBFs are used in this case:

{\small
\begin{subequations}
    \begin{align}
        h_{fc}(\mathbf{x},t)&{=}\bigg\{\begin{array}{ll}
        \Delta x_{fc}{-}(1+\epsilon)v{-}\dfrac{(v_{fc}{-}v)^2}{2a_{l}}& \textrm{if $v{\geq}v_{fc}$}  \\
        \Delta x_{fc}{-}(1+\epsilon)v & \textrm{else}
        \label{eq:hfc-blbr}
        \end{array} \\
        h_{ft}(\mathbf{x},t)&{=}\Bigg\{\begin{array}{ll}
        \Delta x_{ft}{-}\dfrac{(v_{ft}{-}v)^2}{2a_{l}} & \textrm{if $ \Delta x_{ft}{\geq}0$ , $v{\geq} v_{ft}$}  \\
        \Delta x_{ft}& \textrm{if $ \Delta x_{ft}{\geq}0$ , $v{<}v_{ft}$}\\
        \Delta y_{ft}{-}0.1\epsilon & \textrm{else}
        \label{eq:hft-blbr}
        \end{array}\\[3mm]
        h_{bt}(\mathbf{x},t)&{=}\Bigg\{\begin{array}{ll}
        \Delta x_{bt}{-}\dfrac{(v_{bt}{-}v)^2}{2a_{l}}&\textrm{if $\Delta x_{bt}{\geq}0$ , $v_{bt}{\geq} v$}   \\
        \Delta x_{bt}&\textrm{if $\Delta x_{bt}{\geq}0$ , $v_{bt}{<}v$} \\
        \Delta y_{bt}{-}\epsilon & \textrm{else}
        \label{eq:hbt-blbr}
        \end{array} 
    \end{align}
\end{subequations}
}
\begin{remark}
 For the switches between different FSM states, we take the change from \textbf{L} to \textbf{ACC} state as an example to show the continuity of the system's safety under different CBF constraints. We assume that the ego vehicle does a left lane change maneuver and only vehicle $fc$, $ft$ exist. The corresponding safe sets of CBF \eqref{eq:l-fc} and \eqref{eq:l-ft} are called $\mathcal{C}_{fc}$ and $\mathcal{C}_{ft}$, respectively. In \textbf{L} state, the ego vehicle's state $\mathbf{x}$ is in the intersections of these two sets. When the FSM enters \textbf{ACC} state, CBF \eqref{eq:hfc} will be the only hard constraint in the controller, which will build the same safe set as $\mathcal{C}_{ft}$ since vehicle $ft$ becomes the new leading vehicle. In this case, the intersection of sets $\mathcal{C}_{ft}$ and $\mathcal{C}_{fc}$ is the subset of $\mathcal{C}_{ft}$, which meets the proposed safety-based conditions for switches of CBF constraints in Sec. \ref{sec:safety-condition-switch-cbf}. \end{remark} 

Finally, the system's optimal input will be calculated through a quadratic program, where the CLFs and CBFs in different states will be used to construct the constraints,
\begin{subequations} \label{eq:cbf-clf-qp-control-design}
\begin{align}
   u &= \argmin_{\begin{bmatrix}
    u ~ \delta_{v} ~ \delta_{y} ~ \delta_{\psi}
    \end{bmatrix}^T\in \mathbb{R}^{5} } \dfrac{1}{2}u^THu+p_v\delta_v^2+p_{y}\delta_{y}^2+p_{\psi}\delta_{\psi}^2 \\
    \text{s.t.} & \quad L_fV_{j}(\mathbf{x})+L_gV_{j}(\mathbf{x})u\leq -\alpha_{j} V_{j}(\mathbf{x}) +\delta_{j} \\
    &\begin{small}\quad\dfrac{\partial h_k(\mathbf{x},t)}{\partial t}{+}L_f h_k(\mathbf{x},t){+}L_g h_k(\mathbf{x},t) u \geq{-}\gamma_k h_k(\mathbf{x},t)\end{small} 
\end{align}
\end{subequations}
where $j$ can represent subscript $v$, $y$ or $\psi$ and $k$ can denote subscript $fc$, $ft$ or $bt$. $f$ and $g$ are corresponding items in nonlinear affine kinematic bicycle model \eqref{eq:affine-kinematic-model}. According to the FSM's current state, different combinations of CBFs and CLFs will be used for the optimization problem \eqref{eq:cbf-clf-qp-control-design}.
This quadratic program carries a low computational cost, which makes the controller work at a high update frequency.

\begin{remark}
Notice that the barrier functions used in this paper are time-varying functions. When we construct the CBFs constraints in \eqref{eq:cbf-clf-qp-control-design}, we must use surrounding vehicle's time-varying speed and acceleration information to calculate corresponding partial derivatives.
\end{remark}
\section{Results} \label{Sec:Results}
Having introduced the ruled-based safety-critical lane change controller, we will validate our algorithm through numerical simulations in this section. Simulations are also illustrated in a supplement video\footnote{\url{https://youtu.be/icmy9u2a4z4}}. The controller is simulated to work at an update rate of 100Hz and parameters in TABLE \ref{tab:simulation-parameter} will be used.
\begin{table}
    \setlength{\abovecaptionskip}{0.5cm}
    \setlength{\belowcaptionskip}{-0.5cm}
    \centering
    \caption{Parameters for simulation.}
    \begin{tabular}{c c | c c|  c c}\hline
        \multicolumn{2}{c}{\textbf{Vehicle Parameter}} & \multicolumn{4}{c}{\textbf{Hyperparameter }}\\\hline
        Parameter & Value & Parameter & Value& Parameter & Value\\ \hline 
        $l_f$ & 1.11m&\multirow{3}{*}{H}&\multirow{2}{*}{\scriptsize{$\begin{bmatrix} 0.01&0\\0&0 \end{bmatrix}$}}&$\alpha_v$&1.7\\
        $l_r$ & 1.74m&~&~&$\alpha_p$&0.8\\
        $l_{fc}$ & 2.15m&$\epsilon$ &0.5 & $\alpha_s$ & 12\\
        $l_{rc} $& 2.77m&$p_v$ & 0.1 &  $\gamma_{fc}$& 1\\
        $w_{lc}$ & 0.93m&$p_p$ & 15 & $\gamma_{ft}$ & 1\\
        $w_{rc}$&0.93m&$p_s$ & 400& $\gamma_{rt}$ & 1\\\hline
    \end{tabular}
    \label{tab:simulation-parameter}
\end{table}

As discussed in Remark \ref{rem:kinematic-dynamic-model-limitations}, we apply constraints on the system's input to satisfy the requirement of the kinematic bicycle model and physical limitations of the ego vehicle, which are shown in TABLE \ref{tab:simulation-constraint}.
\begin{table}
    \centering
    \caption{Values for input constraints}
    \begin{tabular}{c c c c c }\hline
    ~&$\beta$ & $\Dot{\beta}$ & $a$ & $a_y$\\\hline
    minimum & \ang{15} & \ang{15}/s & 0.3$g$&0.3$g$\\
    maximum & -\ang{15} & -\ang{15}/s & -0.3$g$ & -0.3$g$\\\hline
    \end{tabular}
    \label{tab:simulation-constraint}
\end{table}

\subsection{Simulation with Pre-designed Typical Scenarios}
\begin{table}
    \centering
    \caption{Initial positions and speeds of the surrounding vehicle for the three typical scenarios numerical simulations.}
    \begin{tabular}{c c c }\hline
        Simulation & Initial Position $(x_1(0), y_1(0))$ & Constant Speed $v_1(t)$ \\\hline
        1 & (55 m, 1.75 m)&22 m/s\\\hline
        2 & (-15 m, 5.25 m)& 19 m/s\\\hline
        3 & (3 m, 8.75 m)& 33 m/s\\\hline
    \end{tabular}
    \label{tab:initial-state-simulation}
\end{table}

To evaluate our controller's performance, we test the algorithm on the ego vehicle in the following pre-designed typical scenarios, which can be used to mimic most driving scenes. For these scenarios, our ego vehicle is simulated to start from initial position $(x(0), y(0)) =$ (0m, 1.75m) with an initial speed $v(0)$ = 27.5m/s and the target lane is the adjacent one on the left. The width of each lane is 3.5m, the ego vehicle's speed limit is $v_l = 33.33$ m/s and its desired speed is set as $v_d = 27.5$ m/s. Another surrounding vehicle is also simulated and it is initialized with a different position and constant speed in each simulation, shown in TABLE \ref{tab:initial-state-simulation}. This surrounding vehicle is set to stay on its original lane for first two simulations and it is required to change to the same target lane as the ego vehicle in the third one.

\begin{figure}
    \centering
    \begin{subfigure}[t]{0.99\linewidth}
        \centering
        \includegraphics[width=8.5cm]{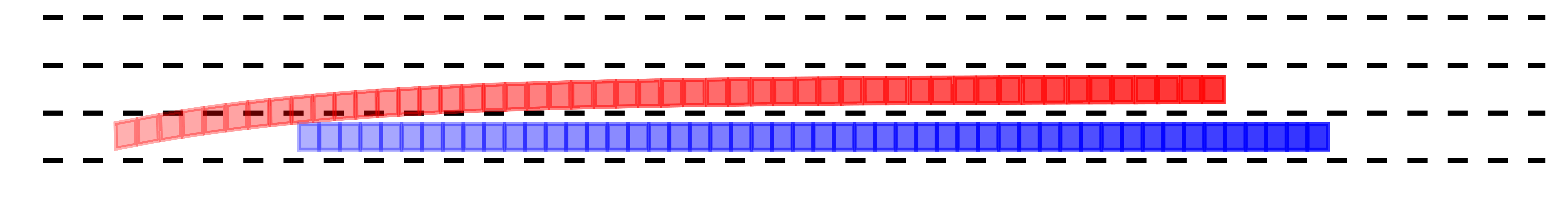}
        \caption{Simulation 1: ego vehicle overtakes a slow leading one.}
        \end{subfigure}
    \begin{subfigure}[t]{0.99\linewidth}
        \centering
        \includegraphics[width=8.5cm]{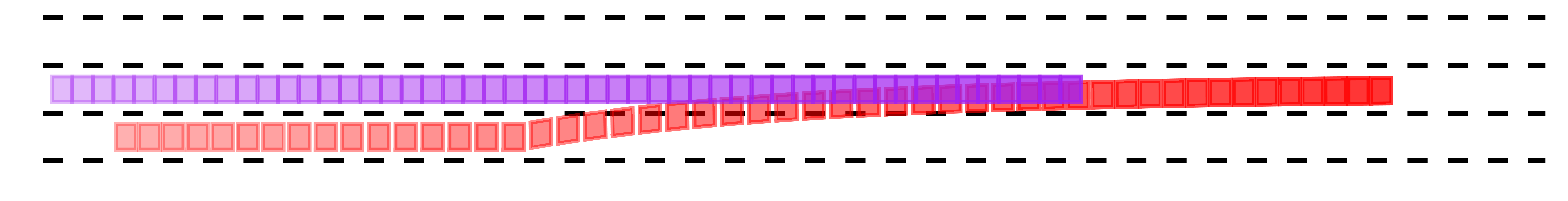}
        \caption{Simulation 2: a target lane's surrounding vehicle approaches ego one.}
    \end{subfigure}
    \begin{subfigure}[t]{0.99\linewidth}
        \centering
        \includegraphics[width=8.5cm]{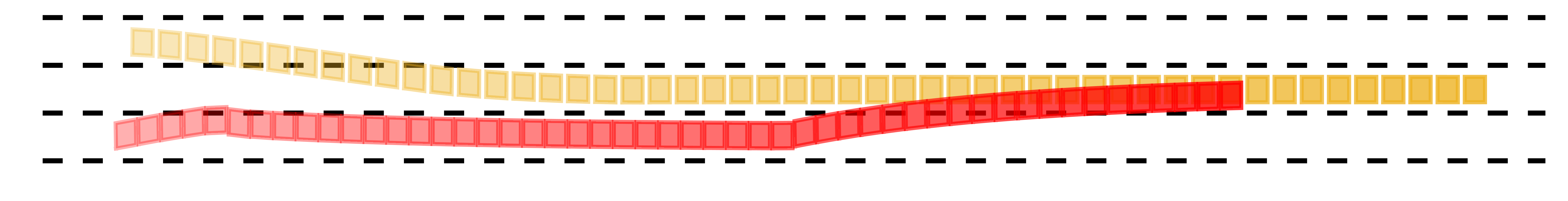}
        \caption{Simulation 3: ego vehicle avoids another surrounding vehicle.}
    \end{subfigure}
    \caption{Snapshots of the ego vehicle and one surrounding vehicle. Red represents the ego vehicle. Blue, purple and yellow show the vehicle $fc$, $bt$ and $ft$, respectively.}
    \label{fig:simulation-trajectory}
\end{figure}

\begin{figure}
    \setlength{\abovecaptionskip}{0.5cm}
    \setlength{\belowcaptionskip}{-0.5cm}
    \centering
    \begin{subfigure}[t]{\linewidth}
        \centering
        \includegraphics[width=\linewidth]{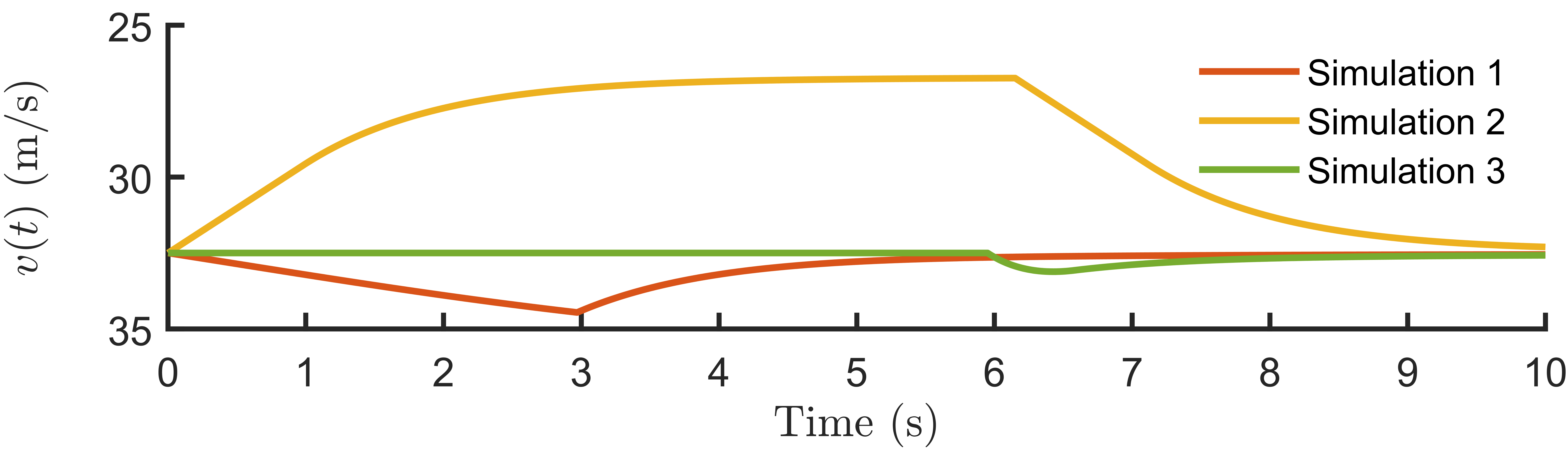}
    \end{subfigure}
    \begin{subfigure}[t]{\linewidth}
        \centering
        \includegraphics[width=\linewidth]{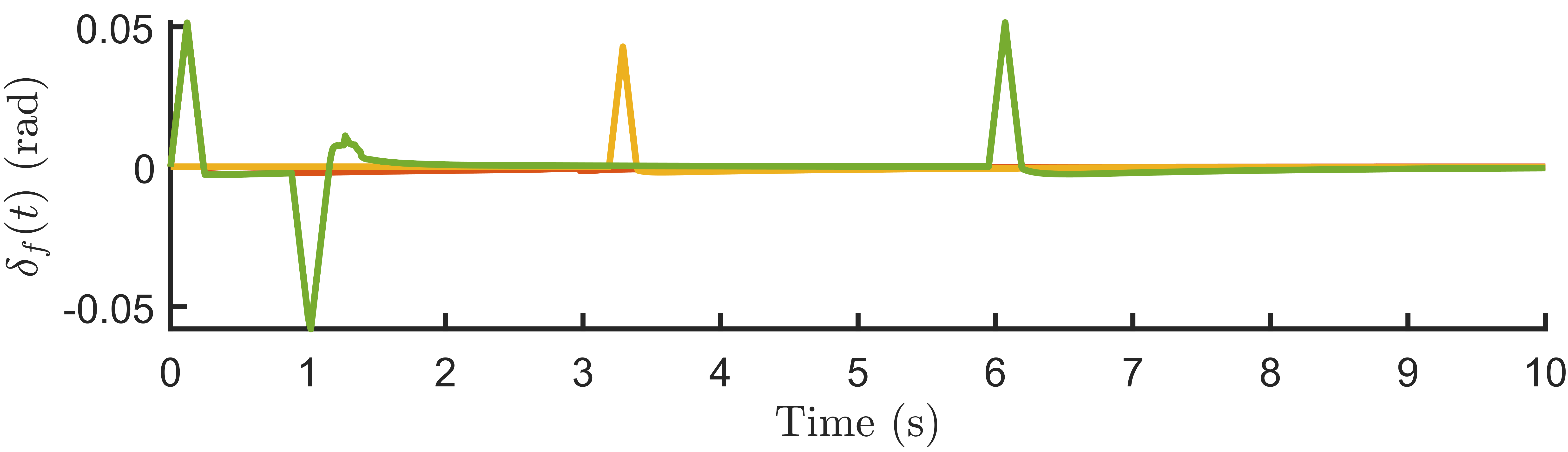}
    \end{subfigure}
    \begin{subfigure}[t]{\linewidth}
        \centering
        \includegraphics[width=\linewidth]{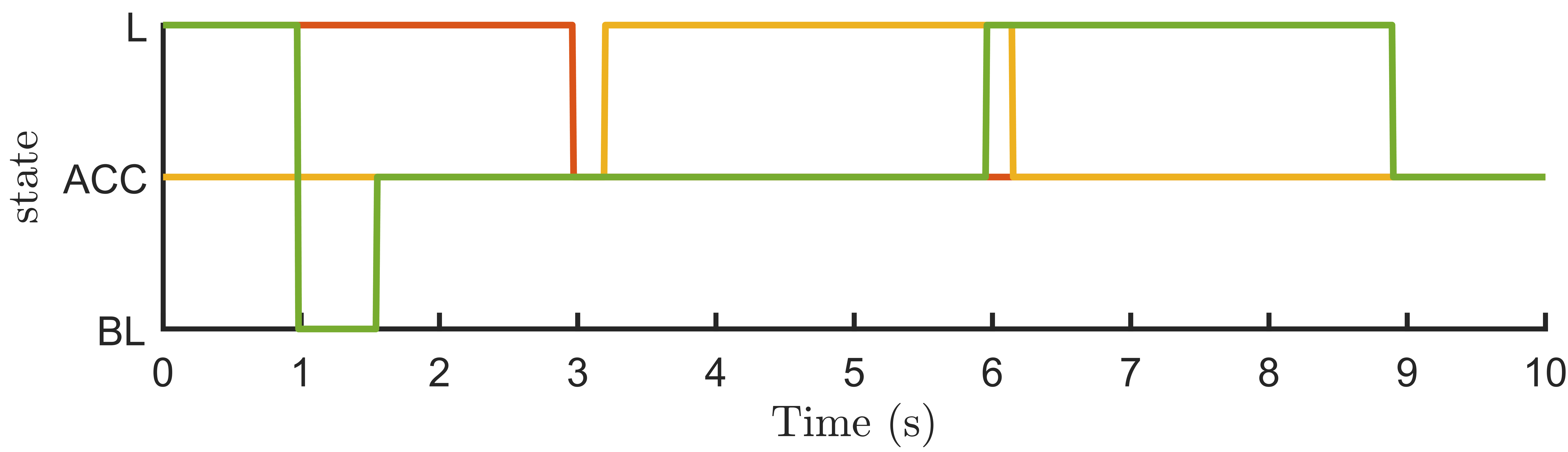}
    \end{subfigure}
    \caption{ The ego vehicle's speed $v(t)$, front steering angle $\delta_f(t)$ and FSM's state during simulations.}
    \label{fig:simulation-control-input}
\end{figure}

The numerical simulations of these scenarios are shown in Fig. \ref{fig:simulation-trajectory} by snapshots. 
The ego vehicle's speed, front steering angle and FSM's state are plotted in Fig. \ref{fig:simulation-control-input}. 

Overtaking a slow leading vehicle occurs frequently and the first simulation is a simple example of this overtaking maneuver. The ego vehicle decelerates to keep away from the slow leading vehicle and changes the lane. Additionally, during the lane change maneuver, the ego vehicle should pay attention to the approaching vehicle in its target lane. Our controller is shown to guarantee the safety of both ego vehicle and this coming vehicle in our second simulation.
The ego vehicle accelerates to gain enough space to change the lane. Finally, a dangerous scenario that may happen during a lane change maneuver is that two vehicles try to change to the same lane at the same time due to lack of observation before changing the lane. In the third simulation, the ego vehicle avoids the other vehicle by driving back to its current lane and then changes to its target lane once it is safe. Our ego vehicle is able to achieve the lane change maneuver and ensure its safety along the trajectory.

\subsection{Simulations with Randomly Generated Tests}
To show statistically the robustness of our control design in different environments, we also deploy tests with numerous driving scenarios with randomly generated surrounding vehicles. We will mimic highway and urban road environments where two groups of random scenarios will be generated. Notice that we adopt the suggested values of speed ranges and lane widths provided by US Federal Highway Administration (FHWA) \cite{stein2007mitigation} to generate two groups of random tests.

\begin{figure*}
    \centering
    \begin{subfigure}[t]{0.99\linewidth}
        \centering
        \includegraphics[width=17cm]{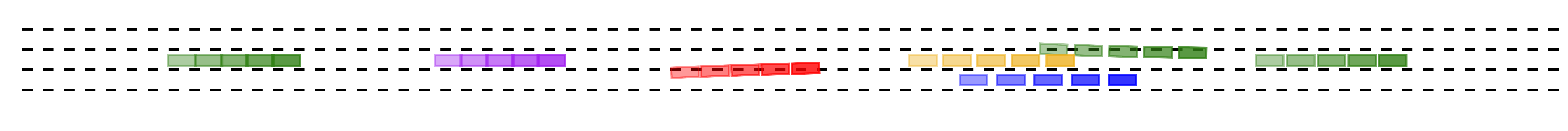}
        \caption{The ego vehicle changes to its target lane in a randomly generated driving scenario successfully.}
    \end{subfigure}
    \begin{subfigure}[t]{0.99\linewidth}
        \centering
        \includegraphics[width=17cm]{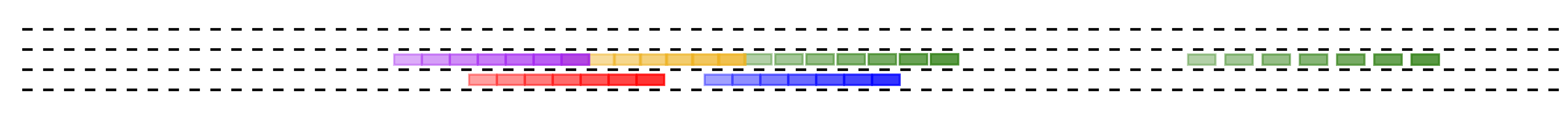}
        \caption{The ego vehicle is still in its current lane after 60s since surrounding vehicles move with similar speed and the distance between them is small.}
        \label{fig:no-change}
    \end{subfigure}
    \begin{subfigure}[t]{0.99\linewidth}
        \centering
        \includegraphics[width=17cm]{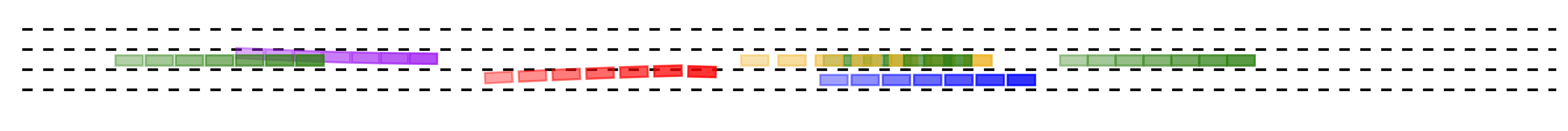}
        \caption{The ego vehicle fails to do a lane change maneuver. A fast vehicle (yellow one) drives ``through" a slow vehicle (one of the green ones). The change of vehicle $ft$ in the controller makes the barrier function $h_{ft}(\mathbf{x},t)$'s value change suddenly. This scenario will not happen in real world.}
        \label{fig:fail}
    \end{subfigure}
   \caption{ Examples of randomly generated scenarios, where red represents ego vehicle; blue, purple and yellow represent vehicle $fc$, $bt$ and $ft$, respectively. Other surrounding vehicles are shown in green.}
    \label{fig:random-test}
\end{figure*}

We can summarize the simulations' basic settings of two groups as follows: only ego vehicle is equipped with our controller and will start from initial position $x(0)=0\text{m}$ and $y(0)$ is equal to the y coordinate of the initial lane's center line in frame $E$ with initial speed $v(0)$, whose value is determined by the driving scenario. The target lane of the ego vehicle is the adjacent one on the left. There will be six vehicles generated randomly around the ego vehicle. One vehicle (denoted as 1st vehicle) is in the initial lane and another four vehicles (denoted as 2nd to 5th vehicle, respectively) are in the ego vehicle's target lane. These five vehicles will move with random initial speed and random acceleration. Finally, one vehicle (denoted as 6th vehicle) is simulated to be in the left adjacent lane of target lane. It will change to the same target lane with a constant speed and begin at a random position. All of above six surrounding vehicles have a vehicle speed lower bound and upper bound, which are determined by the corresponding scenario. Examples of randomly generated scenarios are shown in Fig. \ref{fig:random-test}, where red is the ego vehicle. Random tests' relevant data are summarized in TABLE \ref{tab:random-test-setting}. We do 5000 simulations with a length of 60s for every scenario and a statistical analysis of our controller's performance in urban roads and highway are shown in TABLE \ref{tab:result-random-test}. 

\begin{table}
    \setlength{\abovecaptionskip}{0.1cm}
    \setlength{\belowcaptionskip}{-0.2cm}
    \centering
    \caption{Random test relevant values.}
    \begin{tabular}{c c c}\hline
        Notations & Values & Values\\\hline\hline
        ~&\textbf{Urban Road}&\textbf{Highway}\\
        Driving Lane's Width & 3m&3.6m\\
        Ego Vehicle's $v(0)$ & 13m/s&29m/s\\
        Ego Vehicle's $v_d$&13m/s&29m/s\\
        Ego Vehicle's $v_l$&16.67m/s&33.33m/s\\
        1st vehicle's $x_1(0)$ in $E$ &[25m,40m]&[50m,65m]\\
        2nd-5th vehicle's $x_k(0)$ in $E$&[-50m,50m]&[-85m,85m]\\
        6th vehicle's $x_6(0)$ in $E$&[-50m,50m]&[-85m,85m]\\
        1st-6th vehicle's $v_k(0)$&[11m/s,15m/s]&[26m/s,32m/s]\\
        1st-5th vehicle's $a_k$&[-2m/s$^2$,2m/s$^2$]&[-3m/s$^2$,3m/s$^2$]\\
        Speed lower and upper bound&[10m/s,16.67m/s]&[23m/s,33.33m/s]\\\hline
    \end{tabular}
    \label{tab:random-test-setting}
\end{table}

\begin{table}
    \setlength{\abovecaptionskip}{0.2cm}
    \centering
    \caption{Results of 5000 groups of simulations.}
    \begin{tabular}{c c c }\hline
        Result & In City& On Highway\\\hline\hline
        Change the lane successfully  & 62.46\%&55.58\%\\
        Still in the current lane after 60s&37.06\%&44.22\%\\
        CLF-CBF-QP formulation is not solvable&0.48\%&0.20\%\\\hline
    \end{tabular}
    \label{tab:result-random-test}
\end{table}

From the results, we find that more than 55\% of the simulations finish the lane change maneuver under 60 seconds. Less than 0.5\% of simulations fail to find a solution for the CLF-CBF-QP formulation. In other cases, the ego vehicle does not change to the target lane but meets all safety-critical requirements during the simulations. In order to explore the reasons of later two kinds of results, we look back on details of simulations. The randomly generated vehicles may move at a similar speed with the ego vehicle. Some of them in ego vehicle's target lane have small longitudinal distance with the ego vehicle, which makes it impossible to change the lane (see Fig. \ref{fig:no-change}). Our controller fails when a travelled vehicle, like vehicle $ft$, passes ``through" a slow moving vehicle since we don't constraint the distance between normal vehicles in our simulator (see Fig. \ref{fig:fail}). When this happens, the slow vehicle becomes vehicle $ft$ and the CBF \eqref{eq:hft-blbr} consequently becomes negative, which will make the quadratic program infeasible. Note that vehicles passing ``through" other vehicles will not occur in real life. Additionally, in the real world, drivers will respond to surrounding vehicle's lane change maneuver to ensure the safety and avoid potential threats, for example, decelerate to keep away from cut-in vehicles. This kind of response is not considered in our random test, which makes it more challenging for controller.

\section{Discussion}
\label{Sec:Discussion}
In our proposed algorithm, we use a kinematic bicycle model with small angle assumption on the slip angle for control design. However, since the kinematic bicycle model is based on the zero slip angle assumption, it asks us to limit the ego vehicle's lateral acceleration, which limits the controller's performance. The small angle assumption also causes a mismatch between the real dynamics model and our approximated nonlinear affine dynamics model. Therefore, it is desirable to explore the use of a nonlinear nonaffine dynamic bicycle model in the future work. Besides, in this work, we assume the perception system is able to sense surrounding environment accurately and the controller can receive the sensors' measurements without time delay. To enhance the performance of our safety-critical controller, we need to incorporate a state estimator and introduce a safety margin for the noisy data from distance estimation between the ego vehicle and surrounding ones. The communication time cost between sensors and controller should also be considered in the future work. 

\vspace{-2mm}

\section{Conclusion} 
\label{Sec:Conclusion}
In this paper, we presented a safety-critical lane change control design using rule-based CLF-CBF-QP formulation.
Utilizing a FSM, we divide a lane change maneuver into different states, which allows the proposed strategy to function as both a planner and a controller. A quadratic program based safety-critical control is applied to achieve optimal lane change motion while guaranteeing safety.
We tested the proposed control design using three pre-designed typical driving scenarios and 5000 randomly generated tests in different scenarios to show the effectiveness and robustness.
Experimental results are envisaged for the future.

\balance
\bibliographystyle{IEEEtran}
\bibliography{references}{}
\end{document}